\documentclass[11pt]{article}
\usepackage[utf8]{inputenc}
\usepackage{amsmath, amssymb, amsfonts, tikz, hyperref, comment, url, todonotes, wasysym, import, geometry, amsthm, mathtools, subfig}
\usepackage[
backend=biber,
style=alphabetic,
]{biblatex}
\usepackage{url}
\usepackage[capitalise]{cleveref}
\usepackage[export]{adjustbox}
\usepackage{setspace}
\usepackage{indentfirst}
\theoremstyle{definition}
\newtheorem{example}{Example}[section]
\newtheorem{definition}{Definition}[section]
\theoremstyle{remark}
\newtheorem{remark}{Remark}[section]

\begin{document}

\title{A Comparison of Metrics For The Identification of Partisan Gerrymandering}
\author{Karthik Seetharaman}
\maketitle

\begin{center}
    \begin{abstract}
        Currently, there is currently no effective, standardized way to identify the presence of partisan gerrymandering.
A relatively newly proposed method of identification is ensemble analysis. This is done by generating a large neutral ensemble of voting plans, then comparing the given voting plan against the ensemble through the use of a mathematical metric. In this paper, we survey seven of the most common mathematical metrics used for this identification process, which are the efficiency gap, weighted efficiency gap with weight 2, relative efficiency gap with weight 1, relative efficiency gap with weight 2, mean-median score, partisan bias, and declination. We define and discuss all of these. The analysis is done by performing ensemble analysis on ten elections with each metric to determine whether each metric determines each election as gerrymandered. All ten of these elections have been studied in the literature, so we compare those results to the results found by the metrics to assess accuracy. We show that, while the mean-median score and partisan bias are the most applicable across all seven metrics, they are also the most inaccurate. No significant difference was detected between the remaining five metrics, so our final recommendation is to use one of the four variants of the efficiency gap or the declination if applicable, and the mean-median score if not. The partisan bias was the least accurate of all the metrics tested.
    \end{abstract}
\end{center}

\section{Introduction}

Currently, only $2\%$ of Americans believe that elections are completely fair, and a large reason for this is partisan gerrymandering \cite{johnson_2019}. Partisan gerrymandering occurs when the division of a state into congressional districts for the House of Representatives election purposely favors one party. In Rucho v. Common Cause (2019) \cite{rucho}, the Supreme Court ruled that, although partisan gerrymandering is certainly unfair to the party being gerrymandered against, it is an issue beyond the reach of federal courts. 

In general, it is difficult to identify exactly when partisan gerrymandering is occurring for a variety of reasons. In particular, a state generally being skewed towards one party is not grounds for partisan gerrymandering. The state could have more voters who vote one party than another. Voter distribution, or where voters live, also matters. If voters of one party are typically concentrated in cities and urban areas while voters of the other party live in more rural areas, this could affect the division of the state into congressional districts. 

To help combat this difficulty, many mathematical methods have been created to identify when partisan gerrymandering is happening. These methods are described more in Section~\ref{prelim}, but the main idea is to compare the enacted voting plan to a large space of randomly generated legal distrcting plans (a \textit{neutral ensemble}). In this way, the enacted voting plan is compared to other potential ones, making the identification of significant deviations easier. This method is called \textit{ensemble analysis}.

While many mathematical methods have been proposed in recent years, and some used to legal success (such as the efficiency gap in Wisconsin), there has been relatively little work done on which methods are the most effective. In \cite{warrington2019comparison}, Warrington attempts to compare the most common mathematical metrics, but does so via a suite of hypothetical elections. The paper was also written before the advent of ensemble analysis and thus does not factor in the generation of neutral ensembles into its comparison. This paper attempts to compare the most common mathematical metrics to each other using real, past elections as well as ensemble analysis to create a more up-to-date comparison of the different methods available.

\section{Preliminaries}\label{prelim}

\subsection{The Election Process}\label{process}

We first describe the election process itself. The act of partisan gerrymandering is mainly utilized in elections for the House of Representatives, so those are the elections we will focus on. There are a total of 435 delegates in the House of Representatives, and each state is allotted a number of representatives in the House based on its population compared to the rest of the states. For example, as of the 2020 U.S. Census, Texas has 38 representatives, Florida has 28, Massachusetts has 9, and Vermont has 1 \cite{census.gov_2021}. 

\begin{figure}
    \centering
    \includegraphics[width=10cm]{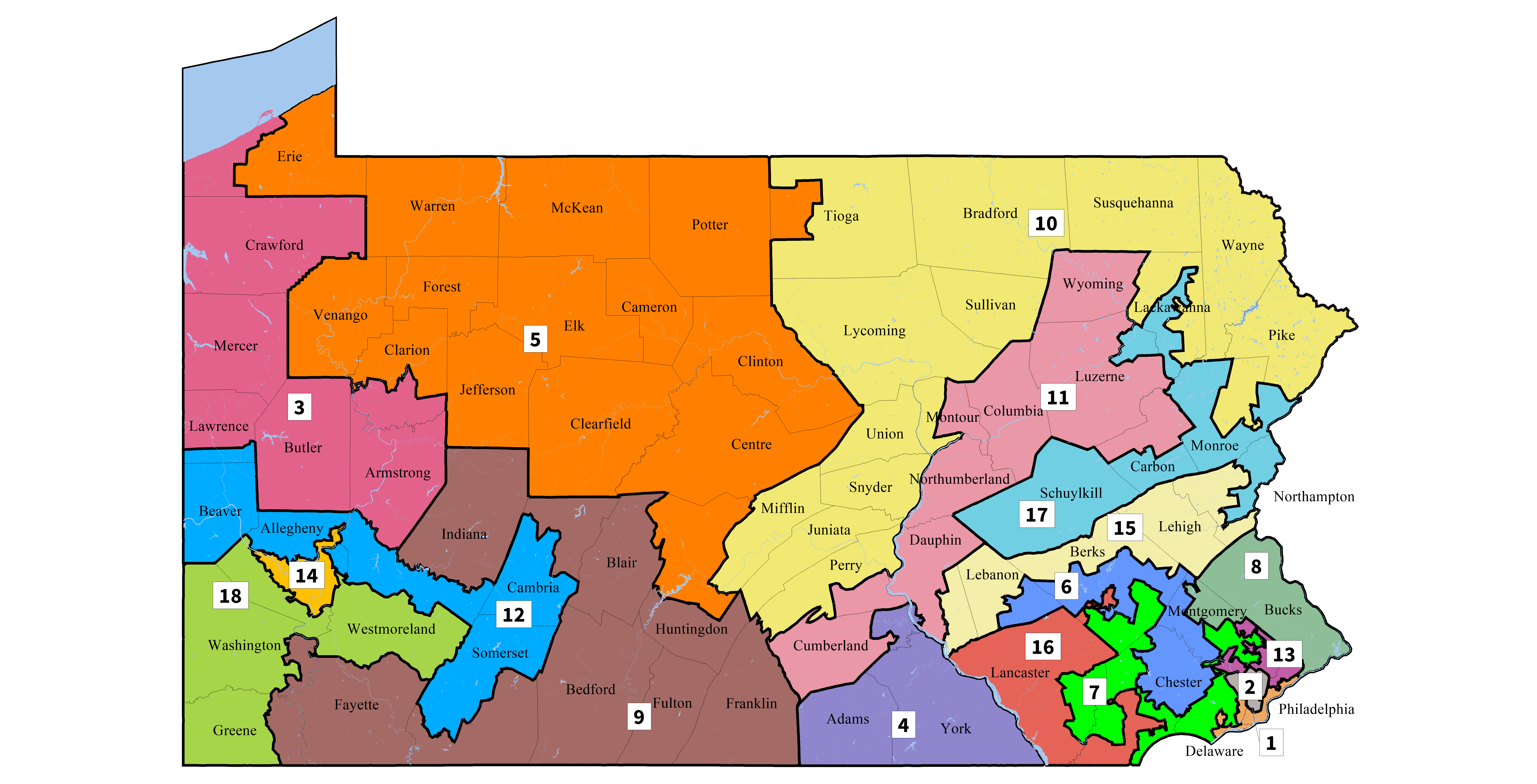}
    \caption{The division of Pennsylvania into 18 congressional districts as of the 2020 U.S. Census. This figure is from \cite{pennsylvania}.}
    \label{fig:pa division}
\end{figure}

The area of each state is then divided into this number of congressional districts, one for each of its representatives. The division of each state into congressional districts is redrawn every ten years during the U.S. Census. Figure~\ref{fig:pa division} shows the division of Pennsylvania into 18 congressional districts in 2011, one for each of its 18 representatives. This map was so famously gerrymandered it eventually had to be changed prematurely to the 2020 U.S. Census \cite{lai_navratil_2018}.

In the biennial House of Representatives election, each district has one representative running from each party (Democrat, Republican, Independent, etc.), one of which will be elected to the House for two years depending on the majority vote in that district. As such, the way the congressional districts in a state are drawn has a large effect on the final outcome of the election. When the districts are drawn in such a way as to purposely favor one party, it is known as \textit{partisan gerrymandering}. Partisan gerrymandering is heavily unfavorable as it results in skewed seat shares in the House that do not accurately reflect the leanings of the state, leading to inaccurate and biased elections.

\subsection{Ensemble Analysis}\label{ensemble}

The identification of partisan gerrymandering in an election is normally a two-step process \cite{asgari2020assessing}. Say we are given a congressional districting that was enacted in a state. First, a large neutral ensemble of plans is generated. This is a large set of congressional districtings (at least 1000, and usually more depending on the size of the state in question) that all could legally be enacted. This usually means that all districts have roughly equal populations and that all districts are contiguous. (In practice, roughly equal means within 2-5$\%$). In a sense, this ensemble represents the space of plans that are feasible to be legally enacted. This generation is done via some computer algorithm. In this paper, we will be using the ReCom algorithm, introduced by DeFord, Duchin, and Solomon in 2019. The details of the algorithm are not vital to this paper, but we refer the interested reader to \cite{deford2019recombination} to learn more. 

The second step of the process is to use some mathematical metric to compare the enacted voting plan to those in the ensemble. There are several choices for such a metric, but the main goal of this step is to determine if the enacted plan is statistically an outlier from the enacted plan. If it is, there is evidence of partisan gerrymandering. Else, there is not significant evidence \cite{deford2019recombination}.

\subsection{Mathematical Metrics}

After a neutral ensemble is generated, the given voting plan must be statistically compared to the ensemble to see if it is an outlier or not. If the given plan is an outlier, there is evidence of partisan gerrymandering. There have been many metrics proposed to perform this statistical comparison \cite{tapp2017relative}, \cite{warrington2018introduction}, \cite{stephanopoulos2015partisan}, \cite{tapp2019measuring}. This paper surveys seven of the most widely used. All of the metrics surveyed in this paper are based solely on the voting data on the election. Other metrics exist to attempt to detect gerrymandering, such as those which measure the compactness of the districts in a given voting plan, but we do not consider those in this paper. 

In what follows, let $V=\{V_1,V_2,\ldots,V_n\}$ be the set of districts of a given state during a given election. For each $1 \leq i \leq n$, let $D_i$ be the number of Democratic votes in the district $V_i$, and $R_i$ the number of Republican votes. We assume two parties, so that $D_i+R_i$ is the total number of votes in district $V_i$ for all $1 \leq i \leq n$. 

\subsubsection{The Efficiency Gap And Variants}

\begin{remark}\label{thirdparty}
All metrics tested in this paper only use votes from two parties, namely the Democratic party and Republican party. Any third-party votes in an election are thrown out when using these metrics. 
\end{remark}

A very popular metric based solely on voting data is the efficiency gap, which is defined in \cite{stephanopoulos2015partisan}:

\begin{definition}\label{standard waste}
A \textit{wasted vote} is a vote for the losing party or a vote for the winning party that is above the $50\%$ majority required to win. In particular, we differentiate between wasted votes for the Democratic party and wasted votes for the Republican party. In a district $i$, we let $WD_i$ be the number of wasted Democratic votes and $WR_i$ be the number of wasted Republican votes. 
\end{definition}

\begin{example}
Take a district $i$ where the Democratic party wins. Then, we have the following two equations: $$WD_i = D_i - \frac{D_i+R_i}{2} = \frac{D_i-R_i}{2}, \qquad WR_i = R_i.$$ The equations are reversed if the Republican party wins. 
\end{example}

\begin{definition}
The \textit{efficiency gap} of an election is defined as $$E_V = \frac{\sum_{i=1}^n (WD_i-WR_i)}{\sum_{i=1}^n (D_i+R_i)},$$ where $V=\{V_1,V_2,\ldots,V_n\}$ is the set of districts of the given state during that election. Higher efficiency gaps point to elections that are more likely to be gerrymandered.
\end{definition}

It is worth understanding the rationale behind the efficiency gap measure, which comes from the two of the main visual characteristics of partisan gerrymandering. These are known as \textit{packing} and \textit{cracking}.

\begin{definition}
\textit{Packing} is when voters of one party are purposefully placed in few districts by the other party, causing them to win those districts by a landslide and lose the others. 
\end{definition}

\begin{definition}
\textit{Cracking} is when voters of one party are purposefully spread across many districts by the other party, causing them to lose most of these districts, often narrowly. 
\end{definition}

\begin{figure}[h!]
    \centering
    \includegraphics[width=10cm]{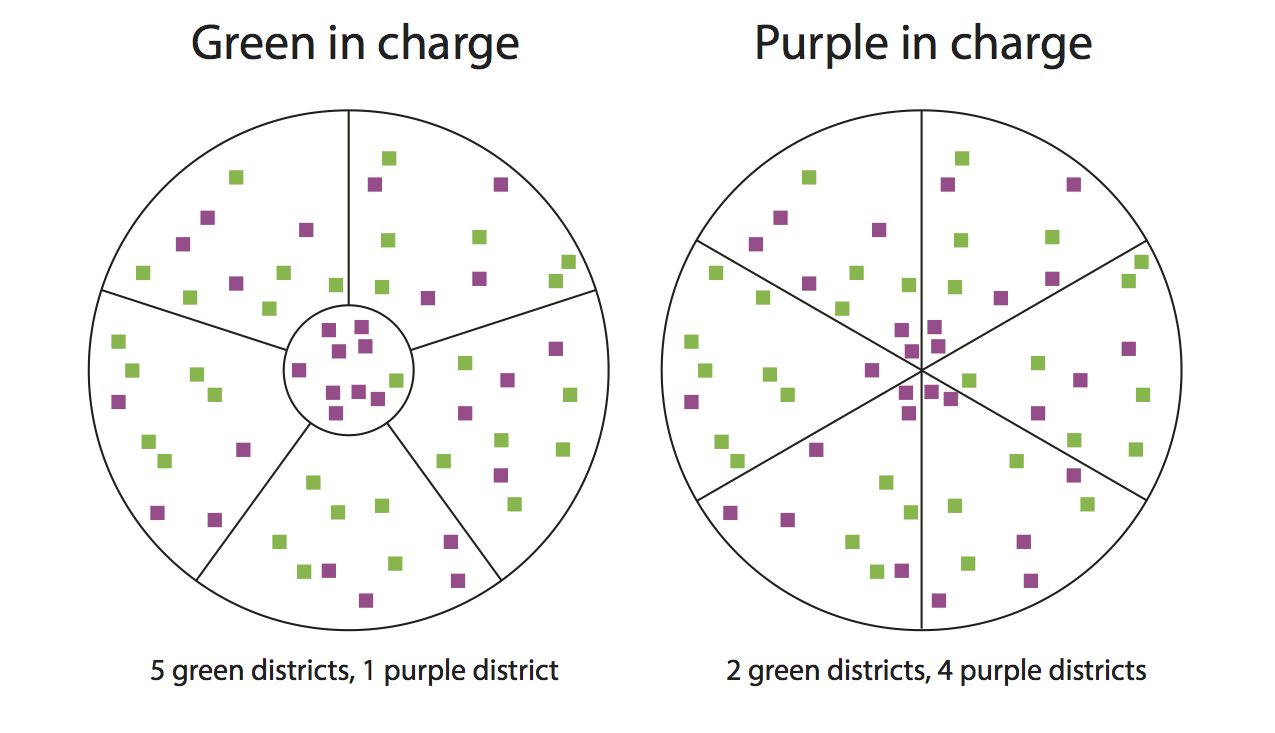}
    \caption{Examples of packing and cracking. The left example shows packing and the right example shows cracking. Notice how the use of such techniques leads to wildly different election outcomes in each case. This figure is courtesy of \cite{princeton}.}
    \label{fig:packing and cracking}
\end{figure}

A good example of the above is shown in Figure~\ref{fig:packing and cracking}. With the same distribution of purple and green voters in the toy state given, different divisions of the state into six districts result in drastically different election outcomes. The left figure shows purple voters packed into the center district, winning that by a landslide and losing the rest narrowly, resulting in a landslide green victory. The right figure shows green voters cracked across many districts, losing them all narrowly to lead to a purple victory. The combination of packing and cracking is one of the most ready visual indicators of partisan gerrymandering, and can have large effects on the outcome of an election. 

The efficiency gap attempts to identify the presence of packing and cracking mathematically by tracking wasted votes. Any district that is narrowly won (a sign of cracking) or won in a landslide (a sign of packing) will have a lot of wasted votes, so elections with a significantly higher number of wasted votes for one party than the other will have a higher efficiency gap, meaning they are more likely to be gerrymandered. 

We also study various variants of the efficiency gap, as described by Tapp in \cite{tapp2019measuring} and \cite{tapp2017relative}.

\begin{definition}
Let $\lambda$ be some fixed positive constant. A \textit{weighted wasted vote} is the same as a wasted vote for votes for the losing party, but votes for the winning party above the majority count for $\lambda$ of a vote rather than one vote. In a district $i$, we let $WWD_i(\lambda)$ be the number of weighted wasted Democratic votes and $WWR_i(\lambda)$ be the number of weighted wasted Republican votes, each with weight $\lambda$. 
\end{definition}

\begin{remark}
To see the rationale behind weighting votes, take the example of $\lambda = 2$. Say a district is won with $70\%$ of the vote. Then, under the normal efficiency gap, $20\%$ of the votes for the winning party were wasted. However, technically, only $30\%$ of the vote was required to win, since the other party received $30\%$. Thus, we should count each wasted vote for the winning party twice, since $40\%$ of their vote was really wasted.
\end{remark}

\begin{example}
Take a district $i$ where the Democratic party wins. Then, we have the following two equations: $$WWD_i(\lambda) = \lambda\left(D_i - \frac{D_i+R_i}{2}\right) = \lambda\left(\frac{D_i-R_i}{2}\right), WWR_i(\lambda) = R_i.$$ If the Republican party wins, the equations are reversed. 
\end{example}

\begin{remark}
Note that the standard wasted votes $WD_i$ and $WR_i$ given in Definition~\ref{standard waste} are equivalent to $WWD_i(1)$ and $WWR_i(1)$, respectively, in the above definition. 
\end{remark}

\begin{definition}
Let $\lambda$ be some fixed positive constant. The \textit{weighted efficiency gap} of an election in a state with district set $V$ is $$WE_V(\lambda) = \frac{\sum_{i=1}^n WWD_i(\lambda) - WWR_i(\lambda)}{\sum_{i=1}^n (D_i+R_i)}.$$
\end{definition}

\begin{remark}
Just as with weighted wasted votes, $WE_V(1) \equiv E_V$. 
\end{remark}

In practice, the only commonly used value of $\lambda$ is $\lambda=2$, so we will only be examining $WE_V(2)$ in this paper. 

Finally, we examine the relative efficiency gap as defined in \cite{tapp2017relative}:

\begin{definition}
Let $\lambda$ be some fixed positive constant. The \textit{relative efficiency gap} of an election in a state with district set $V$ is $$RE_V(\lambda) = \frac{\sum_{i=1}^n WWD_i(\lambda)}{\sum_{i=1}^n D_i} - \frac{\sum_{i=1}^n WWR_i(\lambda)}{\sum_{i=1}^n R_i}.$$ In other words, the proportion of wasted votes is used, rather than the total number of wasted votes.  
\end{definition}

In practice, the only commonly used weights for the relative efficiency gap are $\lambda=1$ and $\lambda=2$, so we will only be examining $RE_V(1)$ and $RE_V(2)$ in this paper. 

\subsubsection{Other Metrics}

We examine three other metrics in this paper, as defined below.

\begin{definition}\label{mean-median}\cite{mcdonald2015unfair}
Let $D$ be the set of all Democratic vote percentages of an election in a state with district set $V$. The \textit{mean-median score} of this election is the difference between the mean and median of $D$.
\end{definition}

\begin{definition}\cite{warrington2019comparison}
As usual, consider an election with district set $V = \{V_1,V_2,\ldots,V_n\}$. Let $M = \frac{\sum_{i=1}^n D_i}{n}$ be the mean Democratic vote percentage, and let $S$ be the set of districts with Democratic vote percentage above $M$. Then, the \textit{partisan bias} of this election is defined as $$PB_V = \frac{|S|}{|V|} - \frac{1}{2}.$$
\end{definition}

The last metric we test, the declination, is more involved to define. Although the metric was coined by Warrington in \cite{warrington2018introduction}, we base our definition off the one given in \cite{campisi2019declination}. Let $V = \{V_1,V_2,\ldots,V_n\}$ be the district set of an election. Let the Democrats win $k$ districts and lose $n-k$ districts. For the declination to be defined, we require $k \geq 1$ and $n-k \geq 1$. Now, consider the set $\{p_1,p_2,\ldots,p_n\}$, the set of Democratic vote shares for each district ordered from least to greatest. Note that we necessarily have $$p_1 \leq p_2 \leq \cdots \leq p_k \leq \frac{1}{2} \leq p_{k+1} \leq \cdots \leq p_n.$$ 

Now, consider two sets $\mathcal{R}$ and $\mathcal{D}$ of points in the coordinate plane: $$\mathcal{R} = \left\{\left(\frac{1}{2n} + (i-1)\frac{1}{n}, p_i\right) : i = 1,2,\ldots,k\right\}$$ $$\mathcal{D} = \left\{\left(\frac{1}{2n} + (i-1)\frac{1}{n}, p_i\right) : i = k+1,k+2,\ldots,n\right\}.$$ In other words, $\mathcal{R}$ is a set of points representing Republican districts, and $\mathcal{D}$ is a set of points representing Democratic districts. Specifically, $\{\mathcal{D} \cup \mathcal{R}\}$ is a set of points equally spaced horizontally whose $y$-value is proportional to the Democratic vote percentage. Points below $y=\frac{1}{2}$ are in $\mathcal{R}$ and points above are in $\mathcal{D}$. 

\begin{figure}
    \centering
    \includegraphics[width=10cm]{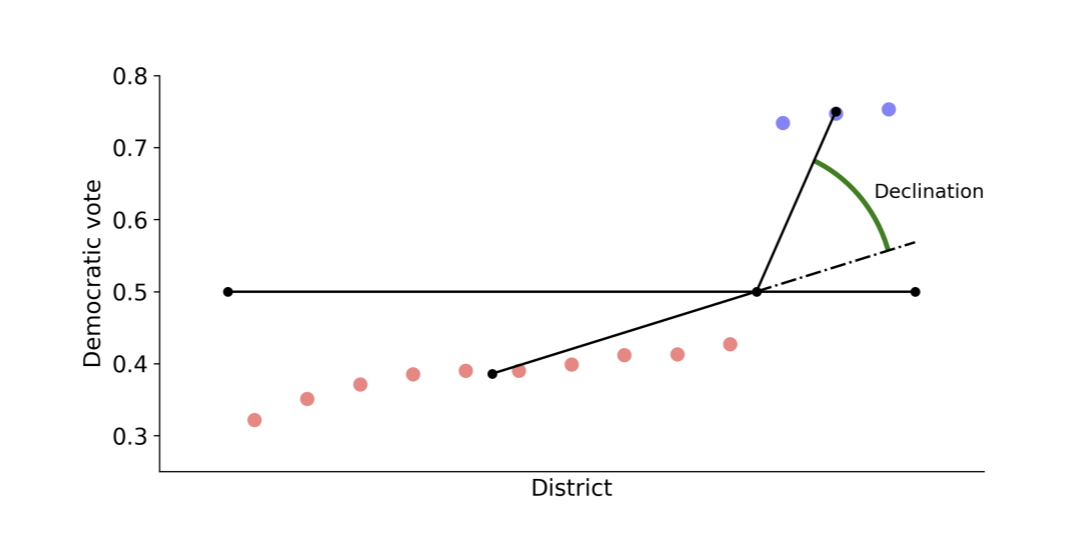}
    \caption{An example calculation of the declination on a Republican-dominated election. Democratic points are colored blue and Republican points are colored red. Figure from \cite{warrington2018introduction}.}
    \label{fig:declinationexample}
\end{figure}

Define $F$ to be the center of mass of $\mathcal{R}$ and $H$ to be the center of mass of $\mathcal{D}$. Also, define $G := (\frac{k}{n}, \frac{1}{2})$. Let $\theta_R$ be the angle $\overline{FG}$ makes with the $x$-axis, and $\theta_D$ be the angle $\overline{HG}$ makes with the $x$-axis. Then, the \textit{declination} is defined as $$\delta = \frac{2}{\pi}(\theta_R-\theta_D),$$ where the $\frac{2}{\pi}$ multiplier is just to ensure the result is in the interval $[-1,1]$. Figure~\ref{fig:declinationexample} shows an example of the declination being calculated on a hypothetical election.

\section{Methods}\label{methods}

\subsection{Materials and Data}\label{materials}

All election data used in this project was gathered from the MGGG Redistricting Lab \cite{github}. Election data is in the form of ``shapefiles," which are files that tabulate the shape of states (as a union of many polygons) as well as election data. Specifically, they show where voters were in relation to their location, which is important for determining if an election was gerrymandered. 

The shapefiles were analyzed using the GerryChain package in Python, an open-source package that allows for the analysis of elections provided in shapefiles \cite{gerrychain}. This package allows the Flip and ReCom algorithms to be run on shapefiles to generate neutral ensembles, and also contains functions for common metrics like the efficiency gap and mean-median score. Slightly less common metrics such as the declination and relative efficiency gap were coded from scratch. 

\begin{figure}
    \centering
    \includegraphics[width=10cm]{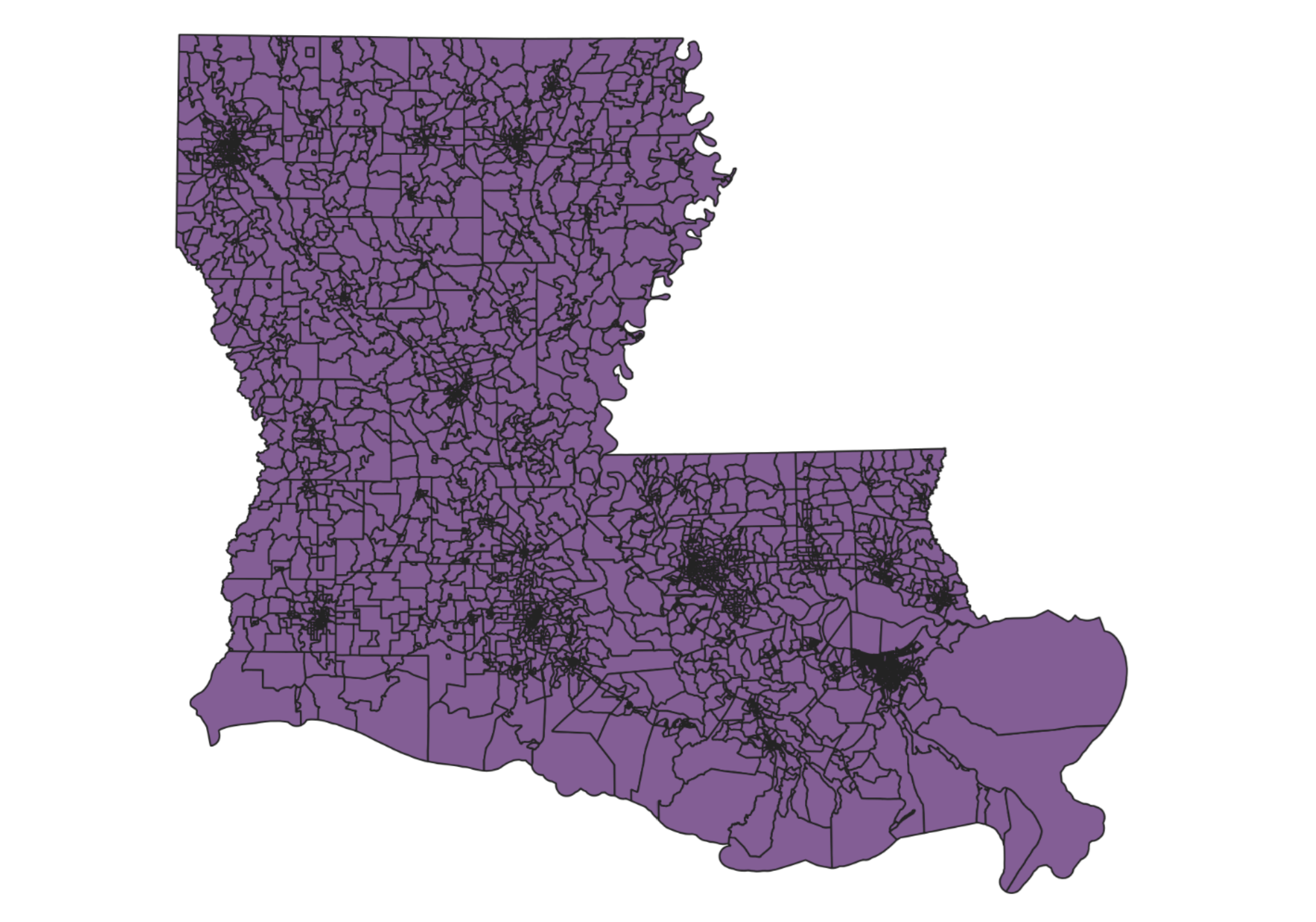}
    \caption{An example of the state of Louisiana being viewed in QGIS. The state is split into several thousand small polygons, corresponding to census blocks.}
    \label{fig:qgis}
\end{figure}

Oftentimes, due to the sheer size of shapefiles (usually in the megabytes), there are often small errors in polygons. These errors must be fixed before the shapefiles can be used in GerryChain, which we did when required through the QGIS software \cite{qgis}. This software allows us to open shapefiles so that to view their polygons visually. Figure~\ref{fig:qgis} shows an example of this. The software also contains methods to fix common problems in shapefiles such as overlapping polygons or irregular shape. 

\subsection{Comparison Methods}\label{methodsstep2}

\begin{figure}
    \centering
    \includegraphics[width=10cm]{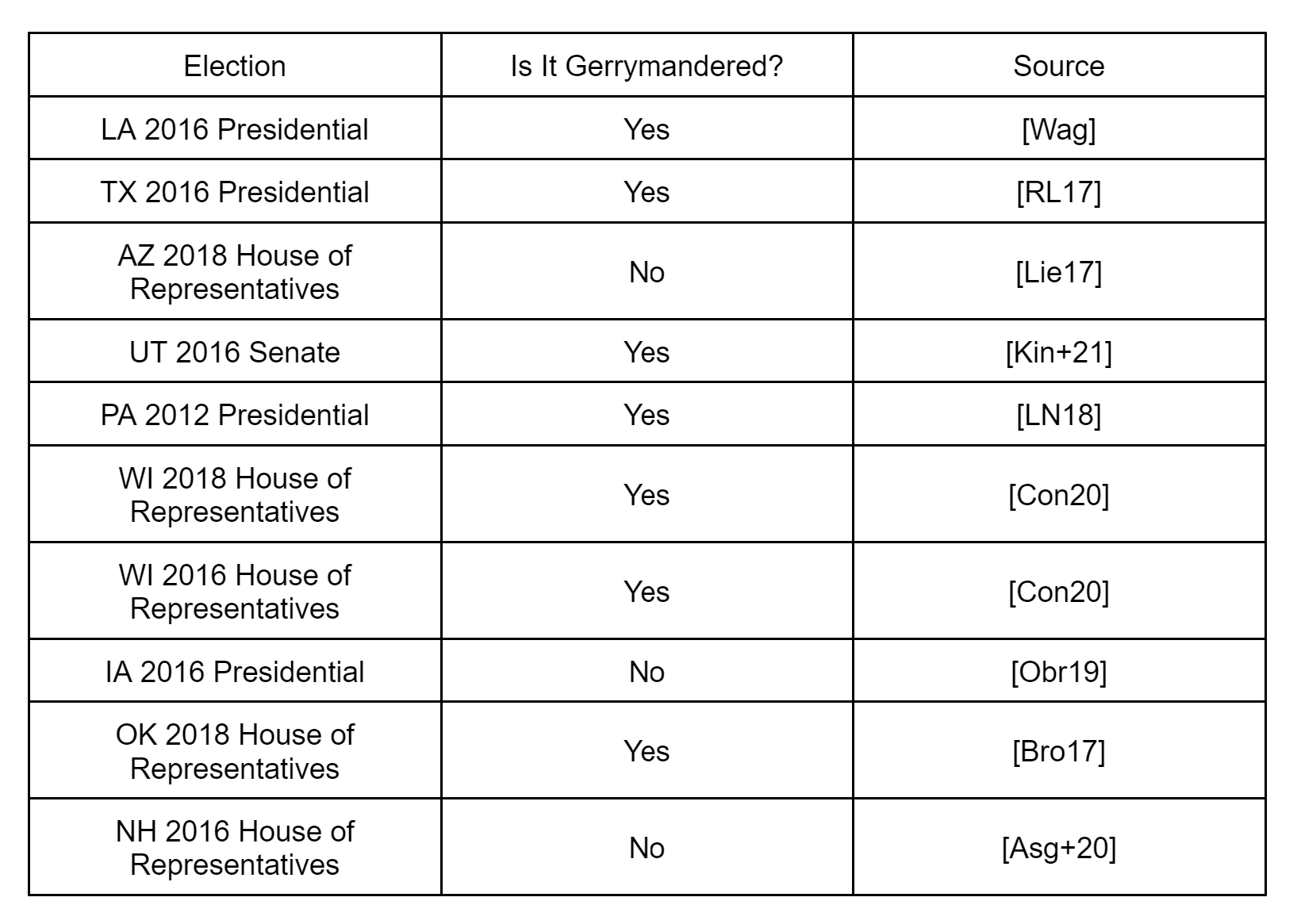}
    \caption{A table showing whether or not each of the ten elections in the chosen sample is gerrymandered, according to past literature. Aside from Pennsylvania, which has been shown in court to be gerrymandered \cite{lai_navratil_2018}, none of these are definitive results, but rather just findings from the previous literature.}
    \label{fig:gerrymandertable}
\end{figure}

For Step 2, we utilize a set of ten studied elections shown in Figure~\ref{fig:gerrymandertable}, gathered as described in Section~\ref{materials}. These elections were chosen to be as varied of a sample as possible. Large states such as Pennsylvania and Texas as well as small states like New Hampshire and Iowa are included. Swing states (e.g. Pennsylvania) and less competitive states (e.g. Oklahoma) were chosen \cite{pew}. 

\begin{remark}
US House elections were used where data was available, but if not, presidential or Senate elections were used instead. All of these elections (House, Senate, Presidential) use the same congressional districting, so which election we use does not matter for the purpose of detecting gerrymandering. 
\end{remark}

For each election $i$, we generate a neutral ensemble $N$ of plans using the ReCom algorithm. As discussed in \cite{deford2019recombination}, we only need at most 10,000 plans to get a sufficiently mixed neutral ensemble. Then, for each metric $j$ being tested, we evaluate all plans in $N$ with $j$. To test whether the enacted voting plan is gerrymandered under metric $j$, we perform a one-sample $t$-test comparing the metric's score on the enacted voting plan to the scores across the ensemble. If the result is statistically significant (using $p<0.01$), the enacted voting plan is said to be gerrymandered under metric $j$. 

Now, to track the accuracy of each metric, we want to be able to track false results, so we need to determine whether or not each election in the sample of ten is gerrymandered. We do this by examining past literature \cite{lieb_2017}, \cite{conklin_2020}, \cite{royden_li_2017}, \cite{king_murri_callahan_russell_jarvis_2021}, \cite{wagner}, \cite{lai_navratil_2018}, \cite{asgari2020assessing}, \cite{brown_2017}, \cite{obradovich_2019}. Figure~\ref{fig:gerrymandertable} lists whether each election was found to be gerrymandered or not based on previous literature.

\section{Results}\label{results}

\begin{figure}
    \centering
    \includegraphics[width=10cm]{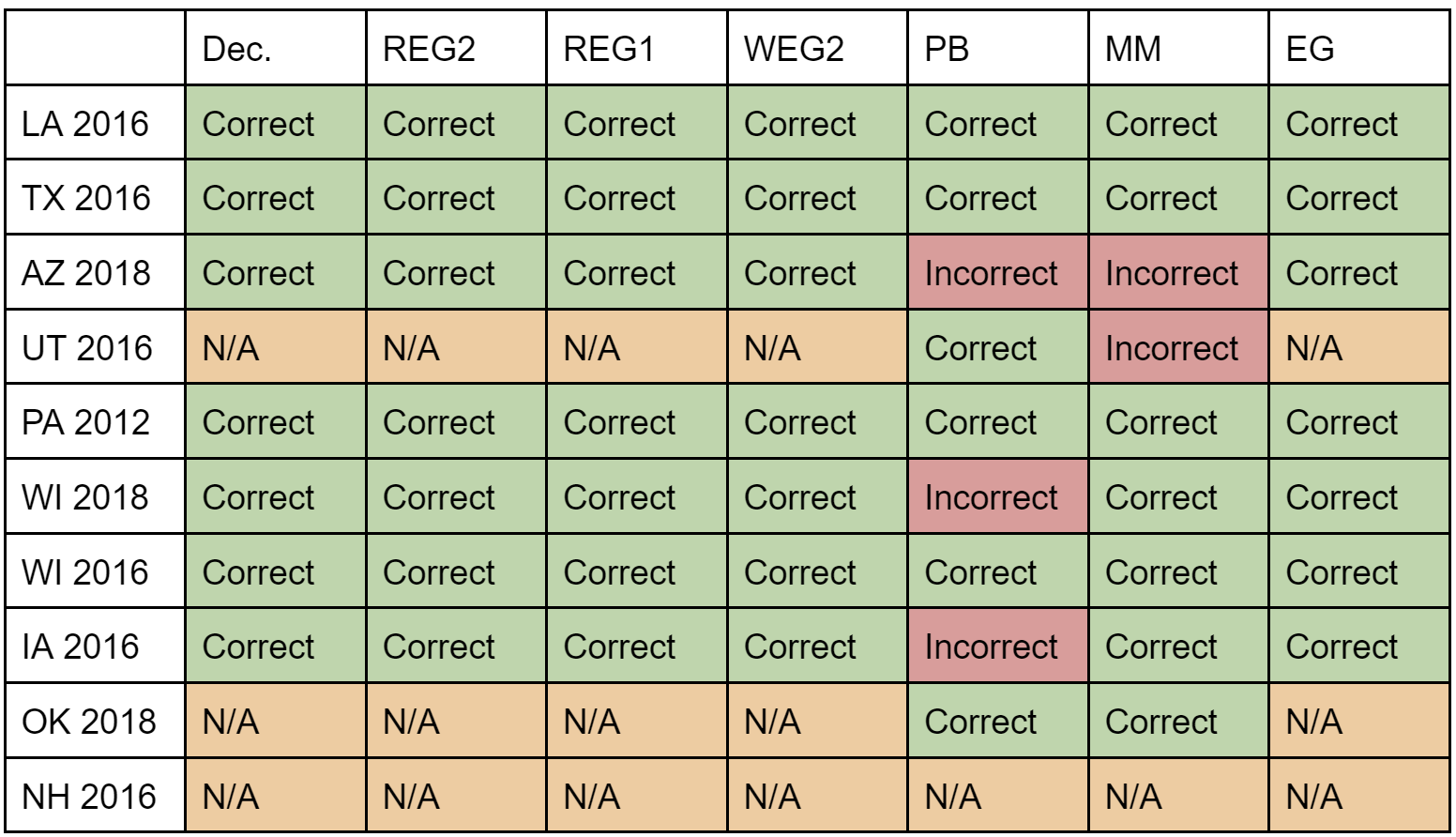}
    \caption{A full table showing, for each pair $(i,j)$ of metric $i$ and election $j$, whether or not metric $i$ returned the correct result, returned the incorrect result, or was inapplicable when applied to election $j$.}
    \label{fig:falseresults}
\end{figure}

Figure~\ref{fig:falseresults} shows the number of false results for each metric out of the ten elections tested, where false results are tabulated as described in Section~\ref{methodsstep2}. Each column represents a metric (from left to right: declination, relative efficiency gap with weight 2, relative efficiency gap with weight 1, weighted efficiency gap with weight 2, partisan bias, mean-median score, and efficiency gap), and each row represents one of the ten elections outlined in Section~\ref{methods}. 

There are three possible classifications for a given pair $(i,j)$, where $i$ is a metric and $j$ is an election. The pair $(i,j)$ is classified as \textit{correct} if the metric correctly determined whether or not the election is gerrymandered based on the classifications given in Figure~\ref{fig:gerrymandertable}. These cells are given with a green background in Figure~\ref{fig:falseresults}. The pair $(i,j)$ is classified as \textit{incorrect} if the metric did not correctly determine whether or not the election was gerrymandered. These cells are given with a red background in Figure~\ref{fig:falseresults}. 

The third classification for a pair $(i,j)$ is \textit{N/A}, when the metric $i$ is not applicable to the election $j$. This can happen for two reasons. Firstly, in elections with two districts, the mean-median score and partisan bias are always 0 on any voting plan, so they are inapplicable to such elections. Secondly, when one party wins all the distrcts in an election, the declination is undefined and all variants of the efficiency gap return the same value, regardless of the actual vote distribution. If the initial election was lopsided, then many plans in the general neutral ensemble will be, causing the metric to not be applicable. These reasons are explained more in-depth in Section~\ref{discussionstep2}. 

As shown in Figure~\ref{fig:falseresults}, the declination, relative efficiency gap with weight 2, relative efficiency gap with weight 1, weighted efficiency gap with weight 2, and efficiency gap all returned 7 correct results, 0 incorrect results, and 3 not applicable results across the 10 elections. The partisan bias metric returned 6 correct results, 3 incorrect results, and 1 not applicable result, and the mean-median score returned 7 correct results, 2 incorrect results, and 1 not applicable result across the 10 elections. In all, there were 48 correct results, 5 incorrect results, and 17 inapplicable results.

Examining the results by election, all metrics determined the state of gerrymandering in LA 2016, TX 2016, PA 2012, and WI 2016 correctly. In WI 2018 and IA 2016, the partisan bias was the only metric to return an incorrect result, while in UT 2016, the mean-median score was the only metric to return an incorrect result. Both metrics returned incorrect results for AZ 2018. In terms of applicability, UT 2016 and OK 2018 had only two of seven metrics (partisan bias and mean-median score) return applicable results, while none of the metrics were applicable to NH 2016. 

\section{Discussion}\label{discussionstep2}

Just looking at the counts of correct, incorrect, and inapplicable results in Figure~\ref{fig:falseresults}, we see that all variants of the efficiency gap and the declination performed the exact same, returning correct results on 7 elections and inapplicable results on the remaining 3. The mean-median score and partisan bias were applicable on 9 elections, but returned incorrect results as well as correct ones. 

We first analyze the inapplicable results. As mentioned in Section~\ref{results}, the elections in Utah, Oklahoma, and New Hampshire were all lopsided, with one party winning all of the districts in the majority of plans generated by the ReCom algorithm. In a lopsided election, the declination is undefined, as at least one Democratic and one Republican district are required for the desired angle to be computable. The efficiency gap is still defined in a lopsided election, but its value turns out to be constant for every lopsided election, not depending on the voter distribution. 

Say, without loss of generality, that an election was lopsided in favor of the Democrats. Then, all Republican votes would be wasted, and the number of wasted Democratic votes would be the total number of Democratic votes minus one-half the total population. In other words, the efficiency gap of a lopsided Democratic election with district set $V=\{V_1,V_2,\ldots,V_n\}$ is just $$\frac{\sum_{i=1}^n D_i - 3\sum_{i=1}^n R_i}{2\sum_{i=1}^n D_i + 2\sum_{i=1}^n R_i}.$$ Notice that this expression does not depend on the distribution of votes across the state at all, but rather just the total number of Democratic votes and total number of Republican votes, which are fixed for any voting plan. Thus, if the majority of the plans generated in a neutral ensemble are lopsided, the efficiency gap does not provide useful information and is thus inapplicable to the election. Similar results hold for all variants of the efficiency gap. 

While the partisan bias and mean-median score were still applicable in Utah and Oklahoma, they were also inapplicable in New Hampshire. This has to do with the fact that New Hampshire is a state with two congressional districts. The mean-median score of an election on $n$ districts is the difference between the mean and median of a set with $n$ elements. If $n=2$, the mean and median will always be equal and the difference will always be 0, meaning the score yields no useful information in this case and is inapplicable. Similarly, the partisan bias of an election on 2 districts is also always 0 since one district will always have Democratic vote percentage above the mean and the other below. 

We are currently unsure exactly why the 5 incorrect results showed up where they did. It is worth noting that three of the five results were in elections that were said to be not gerrymandered (Arizona and Iowa); only three of the ten tested elections were classified as not gerrymandered, so a majority of the false results came from a minority class of elections. Particular attention can be brought to Arizona, which was the only election where both the partisan bias and mean-median score returned incorrect results. Aside from being a non-gerrymandered election, Arizona has another issue: its 7th congressional district registered no Republican voters and $15\%$ third-party voters \cite{washington}. From Remark~\ref{thirdparty}, all metrics throw out third-party voters, so this district would appear fully Democratic in all analyses, which could cause issues. 

From a mathematical perspective, there are differences between the five metrics that had the exact same results on the tested suite of 10 elections. Taking the four variants of the efficiency gap first, Tapp makes the argument that the relative efficiency gap with weight 2 is, mathematically, the best variant of the efficiency gap to use \cite{tapp2019measuring}. One of the reasons for this is due to its behavior in near-lopsided elections, where it is argued that REG2 outputs the most reasonable outputs of the four. In \cite{warrington2019comparison}, Warrington instead mathematically argues for the declination, showing that it increases in absolute value under packing and cracking, two of the main indicators of partisan gerrymandering.

\section{Conclusion}

From Section~\ref{discussionstep2}, it is impossible to pick one mathematical metric as the "best." Overall, the mean-median score and partisan bias are more applicable than the declination and four variants of the efficiency gap, (both were applicable to 9 elections out of the sample versus 7 for the other five), but are less accurate. However, the mean-median score is slightly more accurate than the partisan bias. Thus, we recommend using the mean-median score when the other five metrics are inapplicable, which happens when the election is lopsided or in a very small state (which usually results in a lopsided election). If the mean-median score is also inapplicable, the partisan bias should be used.

Among the four variants of the efficiency gap and the declination, no conclusions can be drawn about their comparison using just the results of this paper, since they all performed exactly the same on every election. Mathematically, there are more arguments in favor of the relative efficiency gap and declination as discussed in Section~\ref{discussionstep2}, but this does not necessarily translate into higher accuracy in practical application. More future work must be done to identify which of these five metrics is the most accurate on past elections, possibly by repeating the methodology of this paper with a much larger sample. 

Another direction of future work is to perform a survey on compactness scores such as the Polsby-Popper score and Reock score \cite{barnes2020gerrymandering}. These are scores that are commonly used to identify partisan gerrymandering by looking at the shapes of districts rather than the voting data itself. These scores are sometimes used in conjunction with voting data metrics like the efficiency gap, and the different combinations of these scores is something worth studying in the future. 

Note that all metrics tested in this paper are far from perfect, being inapplicable in some cases and wrong in others. To remedy this, multiple metrics could be used on an election to attempt to confirm results, or voting data metrics could be combined with the compactness metrics described above. A new metric to identify partisan gerrymandering could also be created that addresses the issues the seven metrics tested here face (namely, lack of support for lopsided or small elections as well as third-party voters).

\section{Acknowledgements}

I would like to thank Dr. Diana Davis of Phillips Exeter Academy for her guidance and assistance throughout the project. I would also like to thank Dr. Kevin Crowthers of the Massachusetts Academy of Math and Science for his mentorship throughout the process. Finally, I would like to thank the PROMYS program, run by Professor Glenn Stevens of Boston University, at which I was introduced to the topic of mathematically identifying gerrymandering.

\printbibliography

\end{document}